\documentclass[prl,showpacs,floatfix,superscriptaddress,twocolumn]{revtex4}
\usepackage{graphicx}
\usepackage{dcolumn}
\begin{document}


\title{Superconducting properties of Fe-based layered superconductor LaO$_{0.9}$F$_{0.1-\delta}$FeAs}
\author{G. F. Chen}
\author{Z. Li}
\author{G. Li}
\author{J. Zhou}
\author{D. Wu}
\author{J. Dong}
\author{W. Z. Hu}
\author{P. Zheng}
\author{Z. J. Chen}
\affiliation{Beijing National Laboratory for Condensed Matter
Physics, Institute of Physics, Chinese Academy of Sciences,
Beijing 100190, China}
\author{H. Q. Yuan}
\affiliation{National High Magnetic Field Laboratory, MS-E536, Los
Alamos National Laboratory, Los Alamos, New Mexico 87545, USA}
\affiliation{Department of Physics, Zhejiang University, Hangzhou
310027, China}

\author{J. Singleton}
\affiliation{National High Magnetic Field Laboratory, MS-E536, Los
Alamos National Laboratory, Los Alamos, New Mexico 87545, USA}

\author{J. L. Luo}
\author{N. L. Wang}
\affiliation{Beijing National Laboratory for Condensed Matter
Physics, Institute of Physics, Chinese Academy of Sciences,
Beijing 100190, China}
%


\begin{abstract}
We have employed a new route to synthesize single phase F-doped
LaOFeAs compound and confirmed the superconductivity above 20 K in
this Fe-based system. We show that the new superconductor has a
rather high upper critical field of over 50 T. A clear signature
of superconducting gap opening below T$_c$ was observed in the
far-infrared reflectance spectra, with
2$\Delta/\textit{k}T_c\approx$3.5-4.2. Furthermore, we show that
the new superconductor has electron-type conducting carriers with
a rather low carrier density.
\end{abstract}


\maketitle

Since the discovery of high-temperature superconductivity in
layered cuprates\cite{Bednorz}, much effort has been made to
explore similar phenomenon in other layered transition metal
oxides. This has led to the discovery of superconductivity in
4d-transition metal ruthenate Sr$_2$RuO$_4$ (T$_c\simeq$1.4
K)\cite{Maeno}, and 3d-transition metal cobaltate
Na$_x$CoO$_2$$\cdot$yH$_2$O (x$<$0.35, y$<$1.3) (T$_c\simeq$4
K)\cite{Takada}. Recently, Kamihara et al. found that the iron- or
nickel-based layered compounds LaOMP (M=Fe, Ni) exhibit
superconductivity with transition temperatures T$_c\simeq$3$\sim$5
K.\cite{Kamihara06,Watanabe} The structure contains alternate
stacking of La$_2$O$_2$ and M$_2$P$_2$ layers along c-axis, with M
ion locating in the center of the P tetrahedron. This discovery
came as a surprise for the scientific community since the
superconductivity comes from the Fe or Ni 3d electrons, while in
usual case the 3d electrons in Fe- or Ni-based compounds tend to
form local moments and to develop magnetic orderings at low
temperature. Very recently, the same group reported that, with the
replacement of P by As and partial substitution of O$^{2-}$ by
F$^-$ in the Fe-based compound to yield La(O$_{1-x}$F$_x$)FeAs,
its T$_c$ could reach as high as 26 K at 5-10 atom $\%$
F-doping.\cite{Kamihara08} This is an exciting event since, except
for the high-temperature superconductivity in copper oxides, the
T$_c$ in this system has already become the highest among layered
transition metal-based compounds. This remarkable discovery not
only opens up new possibilities for exploring novel
superconducting compounds with potentially higher T$_c$, but also
offers opportunity to study the origin of superconductivity from
transition metal d-band electrons, which is expected to shed new
light on the mechanism of high temperature superconductivity in
cuprates.

We have employed a new route to synthesize single phase F-doped
LaOFeAs compound and confirmed the superconductivity above 20 K in
this system. In this work, we present the fundamental
superconducting properties of this system by performing
resistivity, magnetic susceptibility, Hall effect, and optical
measurements. We show that the new superconductor has a rather
high upper critical field of over 50 T. We observed a clear
signature of superconducting gap formation below T$_c$ with an
expected BCS gap amplitude. Besides, the compound has
electron-type conducting carriers with a rather low-carrier
density.

The polycrystalline samples were prepared by the solid state
reaction using LaAs, Fe$_{2}$O$_{3}$, Fe and LaF$_{3}$ as starting
materials. Different from the synthesis method reported by
Kamihara et al. \cite{Kamihara08}, we use Fe$_{2}$O$_{3}$ as a
source of oxygen instead of La$_{2}$O$_{3}$ due to the high
stability of Lanthanum oxide. Lanthanum arsenide (LaAs) was
obtained by reacting La chips and As pieces at 500 $^{o}$C for 12
h then at 850 $^{o}$C for 2h. Mixtures of four components were
ground thoroughly and cold-pressed into pellets. The pellets were
placed into Ta crucible and sealed in quartz tube under argon
atmosphere. They were then annealed for 50 h at a temperature of
1150 $^{o}$C.

\begin{figure}[b]
\includegraphics[width=7cm,clip]{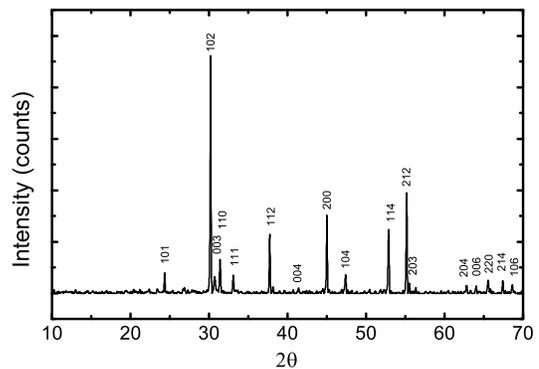}
\caption{X-ray powder diffraction pattern of
LaO$_{0.9}$F$_{0.1-\delta}$FeAs.}
\end{figure}

The phase purity was checked by a powder X-ray diffraction method
using Cu K$\alpha$ radiation at room temperature. As shown in
Fig.1, the powder X-ray diffraction pattern of the resultant is
well indexed on the basis of tetragonal ZrCuSiAs-type structure
with the space group P4/nmm \cite{Quebe}. No obvious foreign phase
was detected, ensuring the proposed treatment was successful to
obtain a single phase LaO$_{0.9}$F$_{0.1-\delta}$FeAs sample. In
the present study we adopt 0.1-$\delta$ for the Fluorine -
concentration in the sample because we can not evaluate possible
evaporation loss of F$^{-}$ component during the high temperature
annealing. The lattice parameters of
LaO$_{0.9}$F$_{0.1-\delta}$FeAs are a= 0.4024 nm and c=0.8717 nm,
respectively, obtained by a least-squares fit to the experimental
data. Compared to the undoped phase LaOFeAs \cite{Quebe}, the unit
cell volume was reduced $\sim$1$\%$ upon $\sim$10$\%$ F-doping,
indicating a successful chemical substitution.

\begin{figure}[b]
\includegraphics[width=7cm,clip]{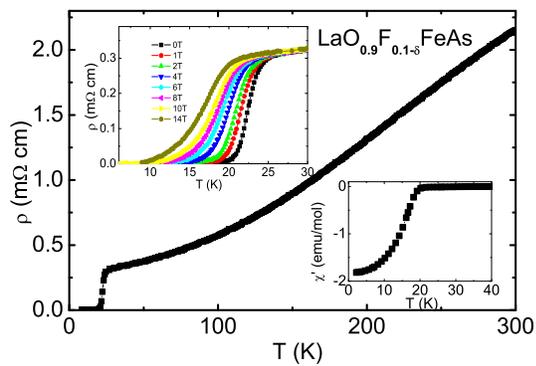}
\caption{(Color online) The electrical resistivity vs temperature
of LaO$_{0.9}$F$_{0.1-\delta}$FeAs. Lower inset: temperature
dependence of the real part of ac magnetic susceptibility. Upper
inset: the resistivity vs temperature curves at selected magnetic
fields.}
\end{figure}

The electrical resistivity was measured by means of a standard
four-probe method in a Quantum Design physical property
measurement system (PPMS). The T-dependence of the resistivity at
zero magnetic field is shown in Fig. 2. With decreasing
temperature, the resistivity decreases monotonously and a rapid
drop was observed starting at about 26 K, indicating the onset of
superconductivity. To confirm the superconductivity, we measured
the ac magnetic susceptibility with a modulation field of 10 Oe
and 333 Hz below 30 K on the same sample. The lower inset of Fig.
2 shows the temperature dependence of the real part of ac magnetic
susceptibility. We clearly observed the appearance of
superconducting diamagnetism at 20 K. The broadening of the
magnetic transition reflects certain inhomogeneity in the
polycrystalline sample.

The upper critical field H$_{c2}$ is one of the important
parameters to characterize superconductivity. To get information
about H$_{c2}$ of LaO$_{0.9}$F$_{0.1-\delta}$FeAs sample, we
measured the electrical resistivity under selected magnetic fields
up to 14 T in PPMS, as shown in the upper inset of Fig. 2. With
increasing the field, the transition temperature T$_{c}$ shifts to
lower temperature and the transition width gradually becomes
broader, similar to the high T$_c$ cuprate superconductors
\cite{Palstra88,Welp89}, suggesting the strong anisotropy of the
critical field, as expected from the two-dimensional electronic
structure\cite{Leb07}. As the T$_c$ was suppressed by only several
kelvins under a field of 14 T, the H$_{c2}(0)$ value is obviously
very high.

It is highly desirable to have more direct information about the
upper critical field. For this purpose, we determined H$_{c2}$, in
pulsed magnetic fields up to 55 T at Los Alamos, using a
tunnel-diode oscillator (TDO) technique in which two small
counter-wound coils form the inductance of a resonant
circuit\cite{Yelland}. The resonant frequency, in our case about
31 MHz, can depend on both the skin-depth (or, in the
superconducting state, the penetration depth) and the differential
magnetic susceptibility of the sample. A piece of thin sample
(about 1.3$\times$0.75$\times$0 .15 mm$^3$) was inserted in the
coils, which were immersed in $^3$He liquid or $^3$He exchange
gas, temperatures being measured with a Cernox thermometer 5 mm
away from the sample.

\begin{figure}
\includegraphics[width=7cm,clip]{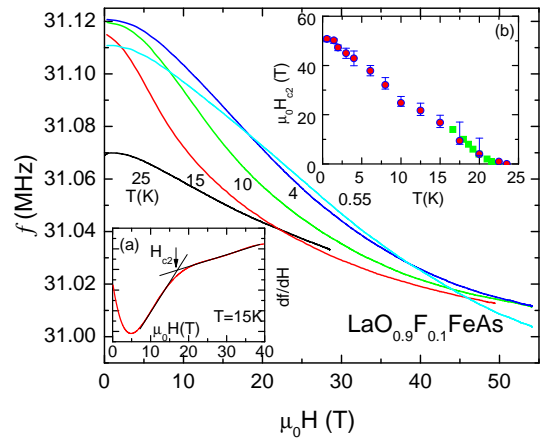}
\caption{(Color online) The field dependence of the TDO
frequencies at T=25, 15, 10, 4, 0.55 K, respectively. Inset (a):
curve for the derivative of TDO frequency with respect to field,
\textit{d}f$/$\textit{d}H, at 15 K. The critical field is
determined by the cross point of the two slopes in the derivative
curve as indicated by the straight lines. Inset (b): critical
field vs temperature phase diagram. The data points obtained from
the mid-point of resistivity drop at T$_c$ are also include (green
square symbol).}
\end{figure}

Figure 3 shows the field dependence of the TDO frequencies at
T=25, 15, 10, 4, 0.55 K, respectively. Superconductivity, with a
relatively broad transition as evidenced in the figure, is
eventually suppressed by applying a high magnetic field. In order
to determine the upper critical fields H$_{c2}$, the TDO
frequencies f(H) were first fitted with suitable polynomial
expressions, then one calculates the derivatives of these fitted
functions with regard to magnetic field. As shown in the inset
(a), the curvature of derivative \textit{d}f$/$\textit{d}H clearly
follows different slopes on the two sides of the critical field
H$_{c2}$, with H$_{c2}$ being determined by the cross point of the
extended slopes as indicated by the straight lines. Inset (b)
shows the extracted critical magnetic field vs temperature phase
diagram. We found that the zero-temperature H$_{c2}(0)$ is over 50
T. In the figure, we also include data points of H$_{c2}$
determined from the resistivity measurement under dc magnetic
field below 14 T, where T$_c$(H) is defined as a temperature at
which the resistivity falls to half of the normal state value
(middle transition). One can see that the two methods give very
consistent results of H$_{c2}$. The H$_{c2}$ vs T curve has a
slight upward curvature.\cite{Remark}

\begin{figure}[t]
\includegraphics[width=7cm,clip]{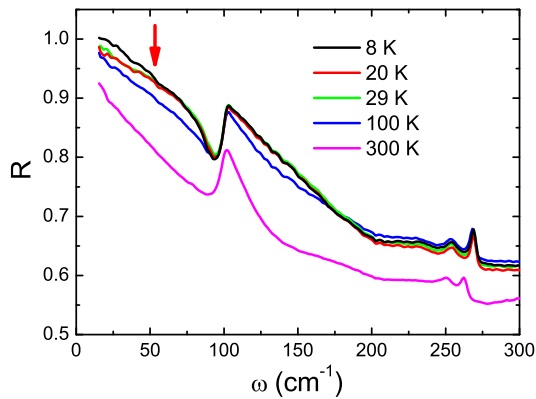}
\caption{(Color online) The far-infrared reflectance spectra at
different temperatures. The arrow indicates the frequency below
which a steep rise appears for the case of T$\ll$T$_c$.}
\end{figure}

The superconducting energy gap is another important parameter for
a superconductor. To obtain the gap information below the
superconducting transition, we measured optical reflectance across
T$_c$ in the far-infrared region on a polished sample in a Bruker
113v spectrometer. A 2.5 mm thick silicon was used as the beam
splitter. The thick silicon beam splitter has superior
long-wavelength performance, enabling us to obtain enough signal
down to very far infrared region. Figure 4 shows the reflectance
curves between 16 and 300 cm$^{-1}$ at different temperatures. The
low frequency reflectance R($\omega$) increases with decreasing
temperature, which is a typical metallic response. A reasonable
estimation for the maximum value of superconducting gap could be
taken from a comparison of the reflectance at T$\ll$T$_c$ and
T$\simeq$T$_c$. Because the sharp transition in magnetic
susceptibility appears at 20 K, little difference could be found
for the reflectance curves at 20 K and 29 K. However, for the
R($\omega$) at 8 K, an abrupt increase below 50$\sim$60 cm$^{-1}$
as indicated by an arrow in the figure could be seen as compared
with curves at 20 and 29 K. This feature is very similar to other
superconducting compounds below T$_c$,\cite{Homes,Ortolani,Lupi}
and apparently could be attributed to the gap formation in the
density of state. The superconducting gap is therefore estimated
to be close to 2$\Delta\simeq$50$\sim$60 cm$^{-1}$. This yields a
2$\Delta/\textit{k}$T$_c$ ratio of 3.5-4.2, being in good
agreement with the expected value within BCS theory.

For a s-wave superconductor without nodes, the reflectance for
T$\ll$T$_c$ approaches to unity abruptly below the gap
energy.\cite{Ortolani} This is not the case for the present
sample, instead R($\omega$) changes in a way very similar to
high-T$_c$ superconductor, for example,
Pr$_{1.85}$Ce$_{0.15}$CuO$_4$.\cite{Homes} This may indicate the
existence of a significant density of states within the gap, or
the pairing symmetry is not s-wave. However, since the sample we
measured is a polycrystalline sample, conclusive information on
the gap symmetry could not be drawn at this stage. For the same
reason, we also would not address other features appeared in the
reflectance curves, such as the pronounced phonon structures, and
the fast decreasing of R($\omega$) with increasing frequency.

To get more information about the conducting carriers, we measured
the Hall coefficient in the normal state using a five-probe
technique in PPMS. Figure 5 shows the Hall coefficient data
between 30 and 200 K. The inset shows the five-leads measurement
configuration, and the verification of the Hall voltage driven by
magnetic field where a linear dependence of the transverse voltage
on the applied magnetic field is observed up to 5 T at 30 K. Two
sets of data were presented in the main panel. The red squares are
Hall coefficient data measured by scanning magnetic field at fixed
temperature, while the solid black curve is R$_H$ determined from
two separate temperature-scan under fixed applied magnetic field
at $\pm$5 T, respectively. They show a rather good match. The
experiment indicates that the Hall voltage is negative, suggesting
electron-type conducting carriers in this compound.

\begin{figure}[t]
\includegraphics[width=7cm,clip]{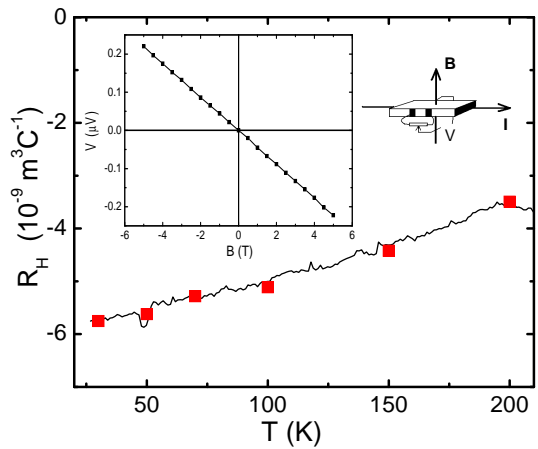}
\caption{(Color online) The Hall coefficient vs temperature for
the sample. The inset shows that the transverse voltage measured
at T=30 K is proportional to the applied magnetic field. The
five-leads measurement configuration is also shown.}
\end{figure}

An interesting result here is that the Hall coefficient is
T-dependent. The absolute R$_H$ value increases with decreasing T.
Large T-dependence of R$_H$ was also observed in high temperature
superconductors, and was regarded as one of the exotic
properties.\cite{Ong} For a metal, the T-dependent R$_H$ is often
explained by the multi-bands effect. When the carrier scattering
rates change with temperature at different rates for different
bands, R$_H$ can become strongly T-dependent.\cite{Ong} Indeed, a
recent band structure calculation for LaOFeP indicates that all
the five Fe d-orbital energy levels are not fully occupied, they
cross the Fermi level E$_F$, leading to five Fermi
surfaces.\cite{Leb07} So it is possible that the T-dependence of
R$_H$ comes from the multi-bands effect. Alternatively, the
T-dependence of R$_H$ could also be caused by a magnetic skew
scattering mechanism: the scattering of conduction electrons from
local moments is asymmetric due to spin-orbital
coupling.\cite{Ong} Magnetic skew scattering has been observed in
various materials with the presence of magnetic moments. As the
compound contains Fe element, there exists chance for the presence
of Fe$^{2+}$ impurities with local moments, then the skew
scattering mechanism might also work here. At present, we could
not distinguish between the above different possibilities. Because
the Hall coefficient changes with temperature, it is not easy to
determine the carrier number precisely from the measurement
result. A rough estimation simply based on the relation R$_H$=1/ne
indicates that the carrier density is rather low, for example, at
200 K, the carrier density is 1.8$\times$10$^{21}$ cm$^{-3}$,
being similar to the high-T$_c$ cuprates.\cite{Ong}

Before conclusion, we would like to address the origin of the
conduction carriers and possible constraint from the experimental
data. From the balance of the valence states in LaOFeAs or LaOFeP,
La$^{3+}$, O$^{2-}$ and P$^{3-}$ are expected to have closed
shells, or to form fully occupied orbitals, only Fe$^{2+}$ have 6
electrons in 3d orbitals. The Fe$^{2+}$ ion locates at the center
of P$^{3-}$ tetrahedral. In such a crystal field, the 3d energy
levels split into the lower e$_g$ (d$_{x^2-y^2}$, d$_{z^2}$) and
the upper t$_{2g}$ (d$_{xy}$, d$_{yz}$, d$_{xz}$). In crystal
structure, the P$^{3-}$ tetrahedral was suppressed a bit along the
c-axis,\cite{Kamihara06} leading to the further splitting of e$_g$
and t$_{2g}$ (the d$_{z^2}$ orbital is pushed to high energy, on
the contrary, the d$_{x^2-y^2}$ and d$_{xy}$ levels are slightly
lowered). According to the band structure calculation for
LaOFeP,\cite{Leb07} all those Fe 3d orbitals are not fully filled,
leading to five Fermi surfaces (without forming local moment), we
can expect that the Hund's rule coupling is rather weak in
comparison with band width broadening in such systems. F$^{-}$
doping into the structure should further increase the electron
number. However, if all the Fe 3d electrons contribute to the
conduction and superconductivity, we would expect a rather high
carrier density. The small carrier density in our measurement
result highly suggests that not all Fe 3d electrons contribute to
the conduction. Apparently, further experimental and theoretical
studies are called on to solve the inconsistency, and to
understand the superconducting mechanism in those compounds.

To summarize, we have made an intensive study of superconducting
properties on F$^{-}$ doped sample
LaO$_{0.9}$F$_{0.1-\delta}$FeAs. We found the onset of
superconductivity occurs close to $\sim$26 K. The main magnetic
transition appears near 20 K. We found rather high upper critical
field of over 50 T and clear signature of superconducting energy
gap opening below T$_c$. A rather low carrier density was revealed
from Hall effect measurement, with the conducting carriers being
of electron-type.

We acknowledge very helpful discussions with T. Xiang and L. Yu.
This work is supported by the National Science Foundation of
China, the Chinese Academy of Sciences, the 973 project of the
Ministry of Science and Technology of China and the PCSIRT of the
Ministry of Education of China. Work at NHMFL is performed under
the auspices of the National Science Foundation, DOE, and the
State of Florida.

\end{document}